\documentclass{article} %

\usepackage{colm2024_conference}

\usepackage{booktabs}
\usepackage{graphicx}
\usepackage{enumitem}
\usepackage{wrapfig}
\usepackage{algorithm}
\usepackage{algpseudocode}
\usepackage[T1]{fontenc}

\usepackage{lmodern}
\usepackage{pifont}

\usepackage{microtype}
\usepackage{amsmath}
\usepackage{colortbl}
\usepackage[utf8]{inputenc}
\definecolor{lightgray}{rgb}{0.9,0.9,0.9}
\usepackage{caption}
\usepackage{subcaption}
\usepackage{xcolor}
\usepackage{setspace}
\usepackage{url}
\usepackage{multirow}
\usepackage{colortbl}
\usepackage{tabularx}
\usepackage{blindtext}
\usepackage{pgfplots}
\pgfplotsset{compat=1.18} 
\usepackage{tikz}
\usetikzlibrary{er,positioning,bayesnet}
\usepackage{makecell}
\usepackage{tipa}
\usepackage{siunitx}
\usepackage{nicefrac}
\usepackage{tocloft}
\usepackage{listings}
\usepackage[raster,skins]{tcolorbox} %
\usepackage{xltabular}
\usepackage{adjustbox}
\usepackage{xurl}
\usepackage{marvosym}
\usepackage{algpseudocode}

\usepackage{amsmath}
\usepackage{amssymb}
\usepackage{mathtools}
\usepackage{amsthm}
\usepackage{enumitem}

\usepackage{algorithm}      % 提供 algorithm 环境
\usepackage{algorithmicx}   % 提供底层算法支持 (关键!)
\usepackage{algpseudocode}  % 提供伪代码格式

\usepackage[table]{xcolor}

\definecolor{outlier1}{RGB}{250,252,255}
\definecolor{outlier2}{RGB}{238,244,250}
\definecolor{outlier3}{RGB}{226,236,245}
\definecolor{outlier4}{RGB}{210,225,240}
\definecolor{outlier5}{RGB}{190,210,230}
\definecolor{outlier6}{RGB}{170,195,220}
\definecolor{outlier7}{RGB}{150,180,210}

\definecolor{loss1}{RGB}{250,253,250}
\definecolor{loss2}{RGB}{242,250,240}
\definecolor{loss3}{RGB}{230,245,230}
\definecolor{loss4}{RGB}{215,235,210}
\definecolor{loss5}{RGB}{200,225,200}
\definecolor{loss6}{RGB}{185,215,185}
\definecolor{loss7}{RGB}{170,205,170}

\definecolor{gap_red1}{RGB}{255,245,205}
\definecolor{gap_red2}{RGB}{255,230,190}
\definecolor{gap_red3}{RGB}{255,210,170}
\definecolor{gap_green1}{RGB}{235,250,205}
\definecolor{gap_green2}{RGB}{220,245,190}
\definecolor{gap_green3}{RGB}{200,235,170}

\author{\textbf{Xiulong Yuan}$^*$, \textbf{Hongqing Chen}$^*$, \textbf{Jiaxuan Peng}$^*$, \textbf{Fan Zhou}$^*$,  \textbf{Zhixiang Ruan}$^*$, \\ \textbf{Zekun Wang}$^*$, \textbf{Bo Zheng}, \textbf{Rui Men}, \textbf{Haiquan Wang}, \textbf{Zhipeng Zhang}, \textbf{Langshi Chen}, \\ \textbf{Man Yuan}, \textbf{Jiaqi Gao}, \textbf{Zhengping Qian}, \textbf{Junyang Lin}, \textbf{Yong Li}$^\dagger$, \textbf{Wei Lin}, \textbf{Junhua Wang}, \textbf{Jingren Zhou} \\
\textbf{Alibaba Group} \\
\small $^*$Equal contribution.\quad $^\dagger$Corresponding authors.
}
\title{Accelerating Compound LLM Training Workloads with Maestro}

\begin{document}

\maketitle

\begin{abstract}
Compound LLM training workloads—such as knowledge distillation and multimodal LLM (MLLM) training—are gaining prominence. These workloads typically comprise heterogeneous components that differ significantly in parameter scale, execution mode (forward-only or full forward-backward pass), and sequence length. Besides, component activation can be data-dependent: in MLLM training, for instance, modality-specific components are activated only when the input contains corresponding modality data, resulting in dynamic computational paths and irregular runtime workloads.
Conventional training frameworks, designed for monolithic models, cannot handle the dual heterogeneity of static configurations and runtime behavior. By enforcing one-size-fits-all training configurations across  components and ignoring input-induced workload variations—they suffer from suboptimal throughput and poor GPU utilization on emerging compound workloads.
In this paper, we introduce Maestro, a \emph{section}-centric training framework systematically addresses the aforementioned dual heterogeneity challenges.
Maestro first reconstructes the compound workload computation into a coarse-grained section graph. Each section is capable of configuring its own tailored parallelism strategy, micro-batch size, and even data-parallel degree. This fine-grained, component-aware mechanism for training configuration and resource allocation effectively addresses the static heterogeneity inherent across different model components.
To tackle data-dependent runtime workload irregularity, Maestro further introduces a wavefront scheduling algorithm that dynamically reorders input samples to orchestrate computation across concurrently active sections while strictly preserving cross-section data dependencies. By doing so, it maximizes inter-section parallelism and minimizes section stalls, thereby improving overall hardware utilization.
% Moreover，to bridge sections with incompatible parallelism schemes, Maestro introduces an asynchronous message queue that performs tensor resharding and communication in the background with minimal overhead.
Maestro has been deployed in production for millions of GPU hours and has demonstrated significant real-world impact: it reduces GPU resource consumption by approximately 40\% on critical workloads—including knowledge distillation and multimodal model training—thereby validating its effectiveness in large-scale, real-world training scenarios.
In this paper, we share Maestro’s design and implementation details to empower the broader research and engineering community in advancing next-generation compound LLM training.

\end{abstract}

\section{Introduction}
The training paradigm for modern large language models (LLMs) is shifting from homogeneous, monolithic Transformer stacks toward compound workloads composed of heterogeneous, functionally specialized components that interact in complex ways~\citep{zhang2025jenga,Xu2025Qwen25OmniTR}.
This shift is particularly evident in the rise of multimodal large language model (MLLM) architectures, which integrate multiple specialized sub-modules—such as modality-specific encoders for feature extraction, an auto-regressive LLM backbone for cross-modal reasoning, and modality-specific decoders for generating outputs across text, image, audio, and other modalities—into a unified training pipeline~\citep{Xu2025Qwen3OmniTR,Team2025LongCatFlashOmniTR,ai2025ming,Bai2026KimiKV}.
Another prominent example of such compound workloads is knowledge distillation, in which a student model is trained to emulate the behavior of one or more teacher models that are often more capable through carefully designed loss functions that facilitate knowledge transfer for efficient model compression. The teacher and student models frequently differ in architecture, parameter size, and execution mode: teachers typically perform inference-only forward passes, while the student undergoes full forward-backward training~\citep{ai2025llama}.

These diverse functional components exhibit markedly different computational, memory, and communication characteristics. Yet, conventional training frameworks typically enforce a uniform global configuration—such as a single parallelism strategy and fixed micro-batch size—across all components. This failure to account for component-level heterogeneity often results in suboptimal training efficiency and underutilized hardware resources~\citep{zhong2025disttrain}.
Moreover, the training recipes in these compound workloads differ significantly from those used for conventional uniformly stacked models. Their unique training strategies—combined with data-dependent computation in sub-modules—introduce dynamic runtime irregularities.
For instance, vision-language model (VLM) training typically leverages a heterogeneous dataset comprising both text-only and text-image samples, aiming to strengthen visual grounding capabilities while maintaining robust performance across general tasks. During training, text-image samples activate the vision encoder for feature extraction, whereas text-only samples bypass it entirely and are routed directly to the LLM backbone. This input-dependent activation pattern yields dynamic computation graphs that vary across batches and thus introduces workload irregularities during runtime~\citep{Wang2025InternVL35AO,bai2025qwen3vltechnicalreport}.
When extended to omni-modal models that accept images, audio, and text as inputs and generate both text and audio outputs, the training data distribution becomes markedly more heterogeneous—spanning text-only, text-image, text-audio, and text-image-audio combinations. This expanded modality spectrum intensifies dynamic runtime irregularities, as the activation patterns of modality-specific encoders and decoders vary substantially across batches, further exacerbating the scheduling challenges posed by input-dependent computation graphs~\citep{Ye2025OmniVinciEA,Xu2025Qwen3OmniTR}.
Optimizing workloads that involve computation of such input-dependent, sparsely activated components lies beyond the scope of existing training frameworks like Megatron-LM, which were primarily designed for monolithic Transformer models. Even with an optimally tuned static configuration, dynamic load imbalance between data-parallel and pipeline-parallel ranks at runtime can readily cause performance degradation and suboptimal resource utilization.

To address the aforementioned challenges, we introduce Maestro, a training framework specifically designed for modern compound LLM training workload that treats heterogeneous components as first-class citizens. Maestro first partitions the original computation graph into \emph{section}s, where each \emph{section} groups sub-modules with similar computational, memory, and communication characteristics. Each \emph{section} can then be assigned tailored training configurations—such as parallelization strategy, micro-batch size, and even data-parallel group size—enabling fine-grained resource allocation and training strategy optimization across heterogeneous model components.
To effectively handle runtime irregularity arising from data-dependent section activation, Maestro incorporates a wavefront scheduler that orchestrates section execution to maximize inter-section concurrency while minimizing data-dependent stalls—prioritizing optimal performance of sections residing on the critical path.
Furthermore, to enable efficient inter-section data transfer with negligible interference to GPU kernel execution, we design and implement an asynchronous, asymmetric message queue based on one-sided remote direct memory access. This mechanism automatically shards and transfers tensors across parallelism domain boundaries with minimal CPU involvement and near-zero GPU overhead.

Experiments on compound workloads—including knowledge distillation and multimodal model training—demonstrate that Maestro achieves up to 1.7× higher training throughput compared to Megatron-LM. Deployed in production for millions of GPU hours, Maestro has accelerated the training of our latest Qwen3.5 model series while reducing GPU resource consumption by approximately 40\%.

\section{Compound LLM Training and Challenges}
The rise of compound LLM training workloads has fundamentally undermined the assumption of workload uniformity that underpins conventional training systems. In this section, we dissect two representative sources of heterogeneity—multimodal model training and knowledge distillation—and analyze how each introduces distinct resource demands and runtime workload irregularities that collectively present critical challenges to training efficiency optimization in existing frameworks.

\subsection{Multimodal model training}
\begin{figure}[!ht]
    \centering
    \includegraphics[width=0.5\linewidth]{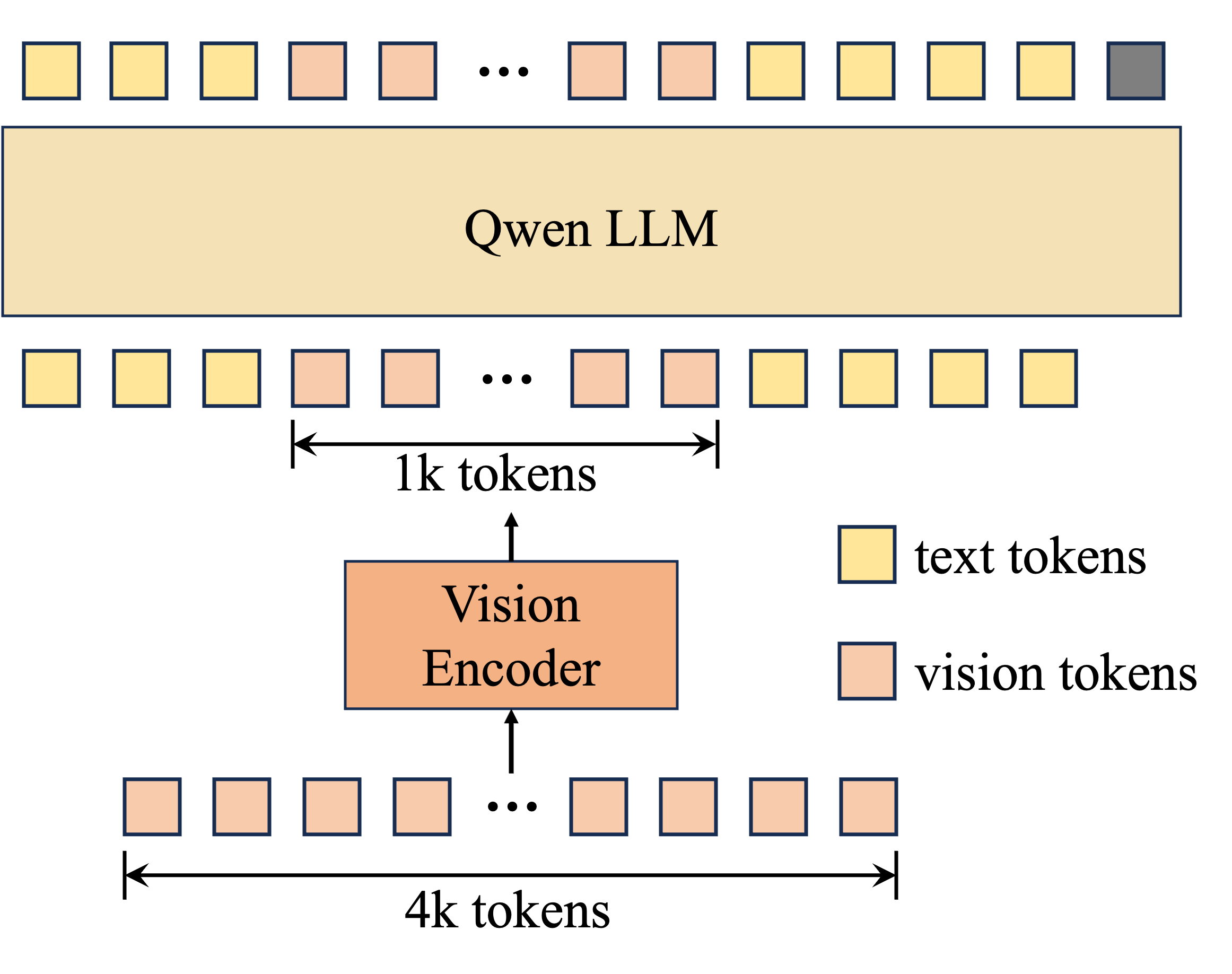}
    \caption{Architecture of Qwen3-VL model.}
    \label{fig:vl_downsample}
\end{figure}

Multimodal large language models (MLLMs) integrate modality-specific encoders and decoders with a shared autoregressive LLM backbone, forming a multi-stage computational pipeline. This pipeline proceeds sequentially through three phases: (1) modality-specific feature extraction by encoders, (2) cross-modal fusion and reasoning within the LLM backbone, and (3) modality-conditioned output generation via dedicated decoders. The resulting workload exhibits pronounced heterogeneity across components—not only in parameter scale but also in computational intensity and memory footprint—posing fundamental challenges for uniform resource allocation strategies.

% The Qwen3-VL-235B-A22B model~\citep{bai2025qwen3vltechnicalreport} exemplifies static heterogeneity: its 0.4B-parameter vision encoder stands in stark contrast to the 235B-parameter LLM backbone~\citep{Qwen3_VL_235B_A22B_Instruct}. Paradoxically, the smaller encoder incurs disproportionate compute cost due to long-sequence attention over raw visual tokens—prior to 4:1 downsampling into compact embeddings. This workload profile renders context parallelism the singularly effective strategy for the encoder, whereas the backbone (as discussed next) demands a hybrid parallelism configuration.
For example, in the Qwen3-VL-235B-A22B model~\citep{bai2025qwen3vltechnicalreport}, the vision encoder contains only 0.4B parameters, whereas the LLM backbone reaches 235B model scale~\citep{Qwen3_VL_235B_A22B_Instruct}. Despite its small parameter count, the vision encoder processes extremely long sequences of visual tokens before applying a downsampling operation (e.g., 4:1) along the sequence dimension to produce compact embeddings for the LLM. This attention computation over long sequences results in disproportionately high computational overhead for end-to-end training relative to its model size. To address this bottleneck, context parallelism is the main necessary parallelism strategy for this component to accelerate attention computation.

% \pjx{Taking the Qwen3-VL-235B-A22B model~\citep{bai2025qwen3vltechnicalreport} as an example, which integrates a vision encoder (0.4B parameters) with the Qwen3 LLM backbone (235B parameters)~\citep{Qwen3_VL_235B_A22B_Instruct}, it is evident that the two components differ drastically not only in parameter scale but also in sequence computation characteristics. The latter stems from an inherent downsampling operation (e.g., 4:1, as shown in Figure~\ref{fig:vl_downsample}), which compresses long visual token sequences into compact embeddings for the LLM, requiring the vision encoder to process input sequences far longer than those handled by the LLM itself. Due to the quadratic time complexity of full self-attention, this results in substantial computational cost for the vision encoder, which accounts for a non-negligible fraction of the total end-to-end training time. Consequently, context parallelism becomes critically necessary for this component. However, this requirement is fundamentally at odds with the combination of tensor, pipeline and expert parallelism typically employed by the LLM decoder, highlighting a critical conflict in parallelism strategies across components.}

In contrast, the LLM backbone confronts a multifaceted optimization challenge. Even when trained on data with moderate sequence lengths, its massive parameter scale imposes severe memory pressure that cannot be alleviated by any single parallelism method. Instead, a hybrid parallelism configuration—orchestrating tensor, pipeline, expert, and context parallelism in concert—is essential to jointly optimize memory footprint, computational density, and communication efficiency. This stands in sharp opposition to the vision encoder's singular reliance on context parallelism, epitomizing the static heterogeneity across components that undermines uniform configuration assumptions in conventional frameworks.
In practice, MLLMs are typically trained on heterogeneous datasets comprising both text-only and multimodal samples~\citep{Wang2025InternVL35AO}. Text-only samples bypass all modality-specific encoders entirely, whereas multimodal samples dynamically activate one or more encoders (e.g., vision or audio encoders) based on input composition. This data-dependent activation pattern introduces runtime irregularity: during multimodal sample processing, the LLM backbone stalls while awaiting modality embeddings; conversely, during text-only processing, modality encoders remain largely idle. When pipeline parallelism is employed, this sample-wise imbalance manifests as dynamic pipeline bubbles that degrade end-to-end throughput. Critically, the degradation scales adversely with pipeline depth: as models grow larger and require deeper pipelines, the relative performance penalty intensifies—leading to disproportionately greater resource wastage despite increased hardware investment ~\citep{Bai2026KimiKV, Team2025LongCatFlashOmniTR,chen2025pipeweaver}.

\subsection{Knowledge Distillation}

\begin{figure}[!ht]
    \centering
    \includegraphics[width=0.4\linewidth]{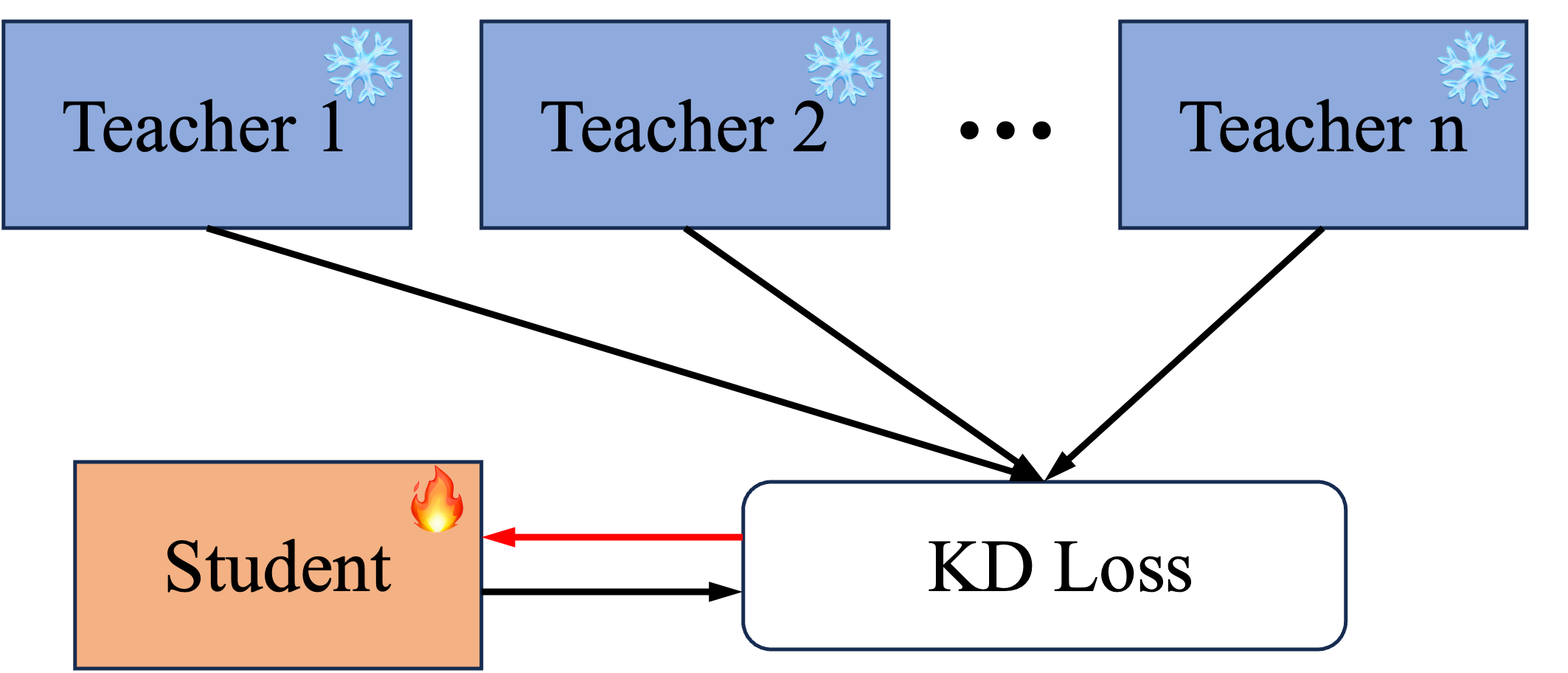}
    \caption{The training workload in knowledge distillation.}
    \label{fig:kd_workload}
\end{figure}

Beyond architectural heterogeneity within a single model, modern training workloads increasingly exhibit cross-model heterogeneity: multiple interacting models—each with distinct execution semantics—are co-scheduled within a single logical training job. Knowledge distillation epitomizes this paradigm, tightly coupling one or more frozen teacher models with a trainable student model in an integrated loop. Critically, these components manifest fundamental execution asymmetry. Teacher models participate exclusively in the forward pass to generate soft targets or logits~\citep{hinton2015distilling, gou2021knowledge}, bypassing backpropagation and parameter updates entirely. In stark contrast, the student model undergoes the full training lifecycle—forward propagation, loss computation, backpropagation, and optimizer-driven parameter updates—bearing the entirety of gradient-related computation and memory pressure.

This execution asymmetry translates into stark disparities in resource requirements and utilization patterns. Although both teacher and student models may demand comparable GPU memory for parameter storage, only the student incurs the full training memory footprint—optimizer states, activations, and gradients—while the teacher, being frozen, requires merely inference-time resources. Consequently, in self-distillation scenarios~\citep{Zhao2026SelfDistilledRO,Hubotter2026ReinforcementLV}, where teacher and student share identical architectures, enforcing a uniform parallelism configuration becomes fundamentally suboptimal. The teacher's lightweight forward-only workload, when subjected to the same aggressive sharding strategy as the student, suffers from excessive device fragmentation—yielding underutilized compute units and unnecessary inter-device communication overhead, despite its minimal computational demands.
Furthermore, increasing the micro-batch size substantially improves the teacher model's throughput with negligible additional memory overhead, while imposing significant activation memory pressure on the student model. Conventional training systems enforce a uniform micro-batch size across all components within a job—dictated solely by the student's memory constraints—forcing the teacher to operate at suboptimal batch scales. This one-size-fits-all policy inevitably leads to underutilized GPU compute resources and degraded end-to-end training efficiency.
\section{Maestro System Design and Implementation}
\label{design}
% 1. overall workflow
% 2. resource allocation
% 3. commuication establish
% 4. 
\begin{figure}[htbp]
\begin{center}
\centerline{\includegraphics[width=0.8\columnwidth]{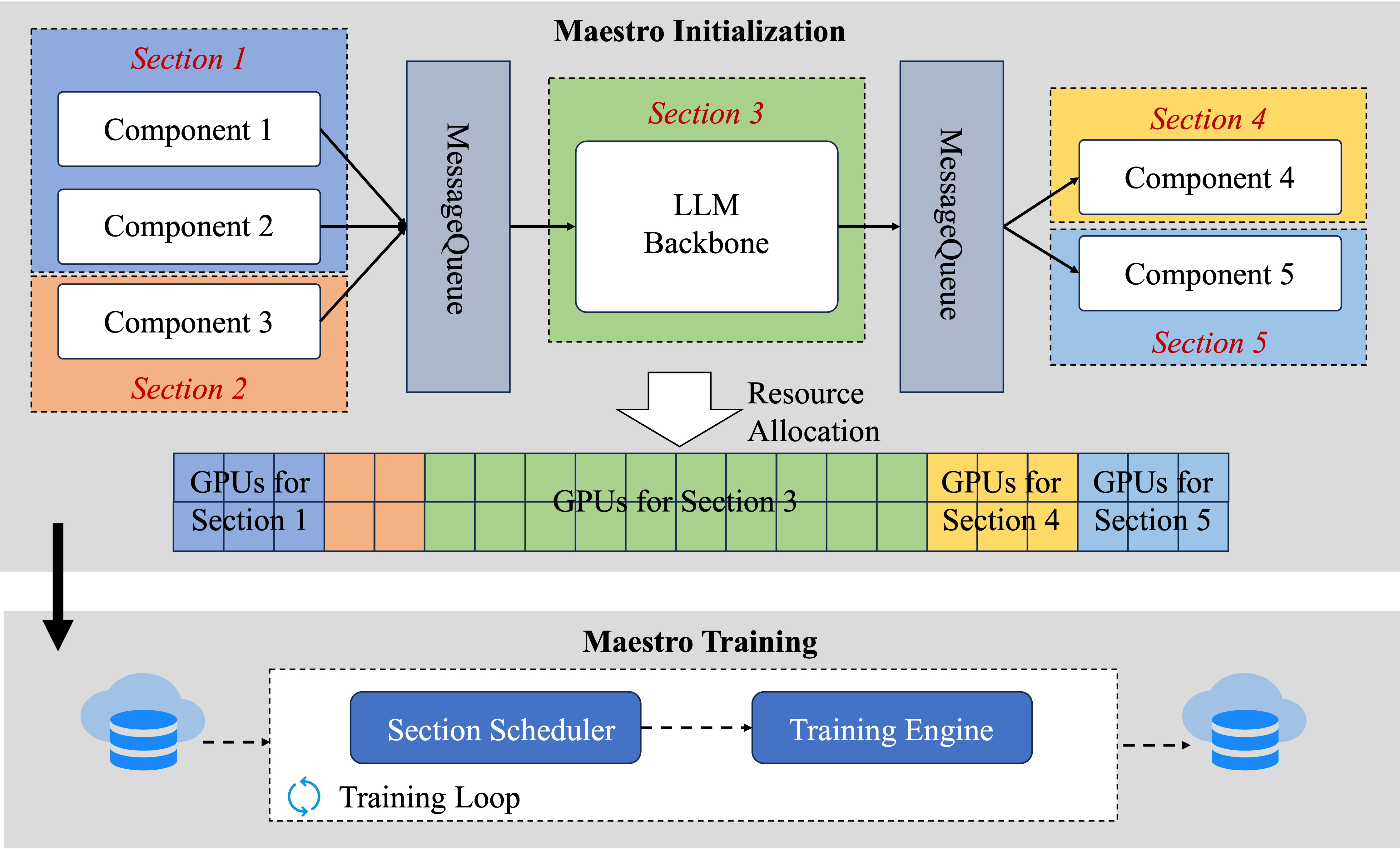}}
\caption{Overview of \textbf{Maestro}.}
\label{fig:maestro_workflow}
\end{center}
\end{figure}

% In compound LLM training, the model is decoupled into logical \textit{sections}, each assigned a dedicated resource allocation. Inter-section data exchange is managed by an asynchronous MessageQueue. During training, a section scheduler orchestrates sample execution employing a sample scheduling algorithm to maximize concurrency and minimize pipeline bubbles across sections.

We illustrate Maestro's overall workflow in Figure~\ref{fig:maestro_workflow}. Given a compound LLM workload—such as MLLM training, Maestro first decomposes models into distinct logical sections based on component boundaries and execution semantics.It then logically partitions the GPU cluster into dedicated resource groups, each sized and shaped according to its section's parallelism strategy (tensor, pipeline, expert, context) and \emph{fan-out} configuration. Each resource group independently initializes its collective communication groups and manages checkpoint loading and saving for its corresponding section.  Crucially, Maestro establishes cross-section communication channels to enable tensor transfer and dynamic resharding across heterogeneous parallelism degrees—particularly between sections employing different tensor parallelism (TP) or context parallelism (CP) degrees. During each training iteration, given a global batch of samples, Maestro leverages an wavefront scheduler that dynamically reorders samples based on their activated sections and exploitable inter-section parallelism. This runtime-aware orchestration simultaneously maximizes cross-section concurrency and suppresses inter-section pipeline bubbles induced by data-dependent activation patterns.

Importantly, Maestro guaranties training equivalence—i.e., it produces identical model updates as the original unmodified training process—and introduces no degradation in model performance.

\subsection{Section Construction}
\begin{figure}[!htbp]
    \centering
    \includegraphics[width=0.5\linewidth]{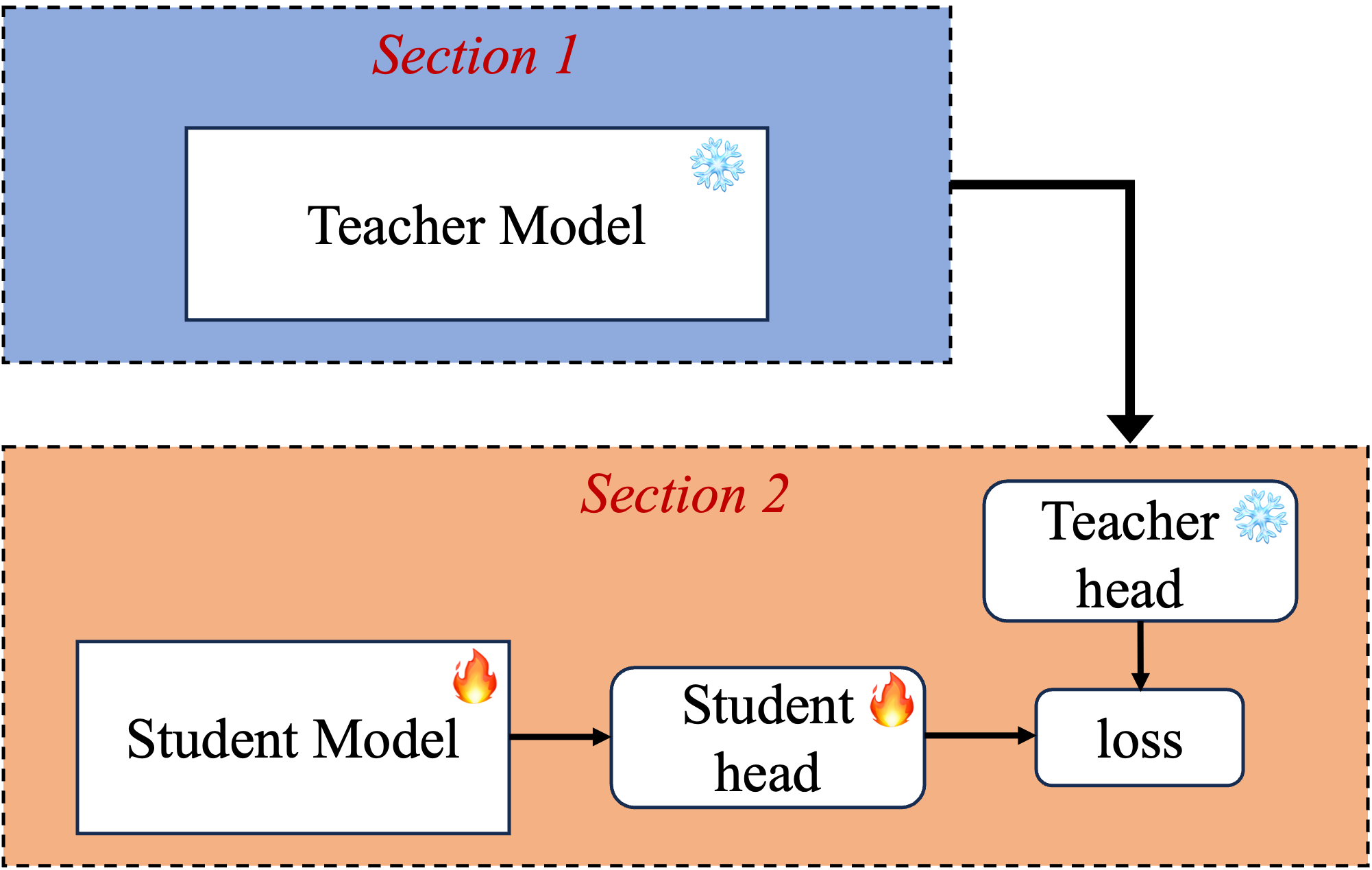}
    \caption{Section construction in knowledge distillation training.}
    \label{fig:distillation_section_construction}
\end{figure}

We first discuss section construction in this part. The section serves as a first-class abstraction for training orchestration in Maestro. In principle, each section corresponds to a logically independent functional component. This not only enables fine-grained, component-aware resource allocation, but also allows sections to execute concurrently to the greatest possible extent. However, certain scenarios demand more sophisticated construction strategies. For instance, in Kullback-Leibler (KL) divergence loss based distillation training~\citep{gou2021knowledge}, the teacher model needs to provide logits tensor to the student model for loss computation. The size of a logits tensor is determined by the vocabulary size and sequence length, whereas the hidden state tensor—used to compute these logits in the final output layer—is governed by the hidden dimension and sequence length. Since the vocabulary size is often an order of magnitude (or more) larger than the hidden dimension, a preferred section construction strategy is to colocate the student model with the final output layers of all teacher models within the same section as shown in Figure \ref{fig:distillation_section_construction}. This way, during execution, the student only needs to fetch the compact hidden states from the teacher models, rather than the much larger logits, thereby significantly alleviating cross-section communication pressure.
Another scenario arises in omni-modal model training. Since samples containing both image and audio modalities typically constitute a relatively small proportion of the dataset, the corresponding encoders are often activated in a mutually exclusive manner during computation. Given that the audio and image encoders have comparable model sizes, they can be co-located within the same section to improve device utilization and minimize resource fragmentation.

%Each section corresponds to one or multiple logically independent functional component—such as one or multiple modality encoders or the LLM—and is designed to execute concurrently with other sections to the greatest extent possible. This enables fine-grained, component-aware resource allocation and maximizes overall training efficiency by overlapping computation across heterogeneous components.
%For example, in 
\subsection{Section Hyper-Parameters Optimization}

After the first decomposition step, each section $s\in\mathcal{S}$ is then assigned a tailored training configuration $\mathcal{C}^s = \{DP^s, TP^s, PP^s, CP^s, \text{mbs}^s\}$, specifying its degrees of data, tensor, pipeline, and context parallelism, along with its micro-batch size. In existing frameworks~\citep{shoeybi2019megatron}, all sections always share the same data-parallel degree. However, this always leads to a fundamental throughput imbalance,  existing among these sections. To illustrate this issue, consider a knowledge distillation workload: the teacher model performs only the forward pass to generate soft labels and thus achieves higher computational efficiency compared to the student model, which executes both forward and backward passes, when both models operate under their respective optimal configurations. Consequently, the faster sections (e.g., the teacher) often suffer from underutilized resources.

\begin{figure}[htbp]
\begin{center}
\centerline{\includegraphics[width=0.5\columnwidth]{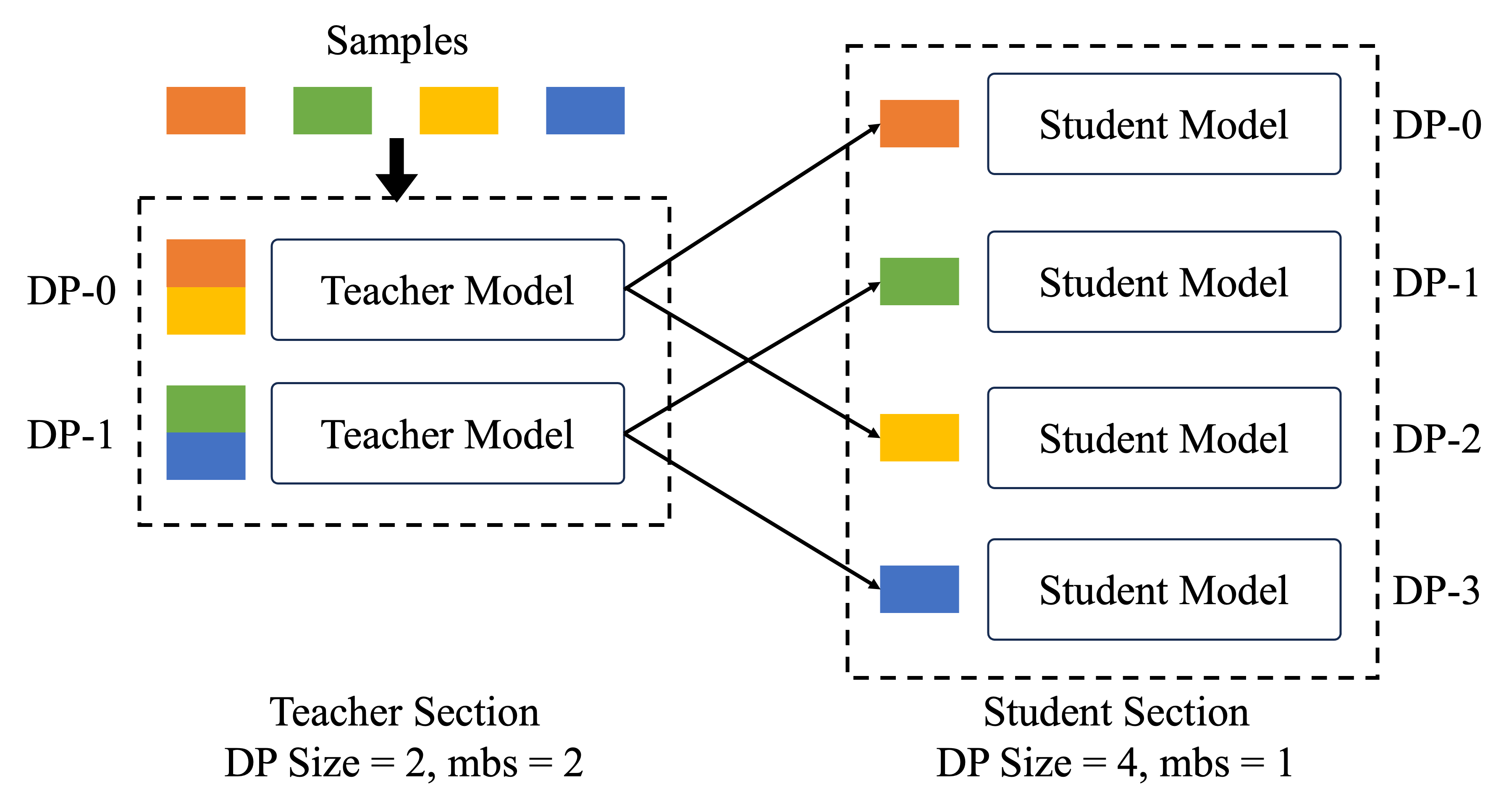}}
\caption{Illustration of fan-out mechanism.}
\label{fig:opd_section_construction}
\end{center}
\end{figure}

To bridge this gap and maximize resource utilization, Maestro introduces a \emph{fan-out} mechanism that decouples their data-parallel execution while preserving correctness. As illustrated in Figure~\ref{fig:opd_section_construction}, each data-parallel rank in the teacher section processes  $ k=2 $  input samples and distributes the resulting outputs to  $ k $  distinct DP ranks in the student section. This allows the student section to consume one sample per DP rank at its natural pace, while the faster teacher section achieves an effective throughput with a lower GPUbudget. The fan-out enforces a precise relationship between their parallelism degrees:
\begin{equation}
    DP^{\texttt{tec}} = \frac{DP^{\texttt{stu}}}{\texttt{fanout}}.
    \label{eq:fanout}
\end{equation}

Therefore, the training configuration turns to be $\mathcal{C}^s=\{DP^s, TP^s, PP^s, CP^s, \text{mbs}^s, \texttt{fanout}^s\}$ based on this mechanism. To maximize overall throughput, the system must allocate resources according to the optimal combination of these per-section configurations. This motivates formulating the problem as an optimization: find  $\{\mathcal{C}^s\}$ that minimizes the end-to-end iteration time  $f(\{\mathcal{C}^s\})$, subject to hardware resource, memory and hyper-parameters constraints:
\begin{itemize}
    \item \emph{Resource constraint:} $ \displaystyle \sum_{s \in \mathcal{S}} N^s \leq N^{\text{GPUs}} $,
    \item \emph{Memory constraint:} $ \displaystyle \max_{g \in \mathcal{G}^s} M_g^s \leq M^{\text{GPU}},  \forall s \in \mathcal{S},$ 
    \item $\texttt{fanout}$ \emph{ constraint:} $DP^{fr} \times \texttt{fanout} = DP^{sr}, \text{where } (fr, sr) \in E \text{ in } G(\mathcal{S},E),$
\end{itemize}
where $N^s$ denotes the number of GPUs allocated to section-$s$; $\mathcal{G}^s$ is the set of GPUs assigned to section-$s$ (with $|\mathcal{G}^s|=N^s$), and $M_g^s$ represents the total memory consumption on GPU $g$ when executing section-$s$, including model parameters, optimizer states, activations, and gradients; $ fr, sr \in \mathcal{S} $  denote two sections connected by a directed data-flow edge in the section dependency graph  $ G = ( \mathcal{S}, E ) $ , with  $ fr $  being the \emph{faster} section (e.g., teacher) and  $ sr $  the \emph{slower} section (e.g., student). Formally:
\begin{equation}
\begin{aligned}
    \min_{\{\mathcal{C}^s\}} \quad & f(\{\mathcal{C}^s\}) \\
    \text{s.t.} \quad 
    & \sum_{s \in \mathcal{S}} N^s \leq N^{\text{GPUs}}, \\
    & \max_{g \in \mathcal{G}^s} M_g^s \leq M^{\text{GPU}}, \quad \forall s \in \mathcal{S}, \\
    & DP^{fr} \times \texttt{fanout} = DP^{sr}, \quad \forall (fr, sr) \in E.
\end{aligned}
\label{eq:conventional}
\end{equation}

However, directly solving Equation~\ref{eq:conventional} is infeasible in practice. Although the valid parallelism degrees for each section are restricted to integer divisors of model-specific structural parameters (e.g., hidden dimension, number of attention heads, and sequence length), the joint configuration space across all sections remains combinatorial due to the multiplicative interaction of per-section choices. Furthermore, every candidate configuration must simultaneously satisfy global resource limits and per-device memory constraints. Given the exponential growth of feasible combinations and the high cost of evaluating each configuration, exhaustive or naive search quickly becomes prohibitively expensive, necessitating a principled and scalable optimization approach.

To overcome these challenges, we start from a key observation: in compound LLM workloads—such as multimodal training or distillation training—system throughput is overwhelmingly bottlenecked by a \emph{critical section}, which incurs the highest computational cost per sample or consumes the largest fraction of end-to-end execution time (e.g., the LLM in vision–language training or the student model in distillation). This critical section effectively defines the \emph{critical path} of the workload: in an ideal scenario where it operates at peak efficiency without stalling, its throughput establishes a fundamental upper bound for the entire pipeline.

Guided by this insight, we adopt a hierarchical optimization strategy that prioritizes the critical path: first, we configure the critical section to approach its theoretical throughput limit; then, we adapt other auxiliary sections to ensure they neither stall nor exert backpressure on the critical path. Specifically, our solution proceeds in two stages:
\begin{enumerate}
    \item \emph{Stage 1 (Critical-first):} Find a near-optimal configuration  $ \mathcal{C}^{\texttt{crit}} $  for the critical section that maximizes its computational efficiency (e.g., MFU), subject to per-GPU memory and hardware constraints.
    \item \emph{Stage 2 (Auxiliary-adaptive):} Given  $ \mathcal{C}^{\texttt{crit}} $  and a chosen \texttt{fanout}, configure each auxiliary section  $ s_{\text{aux}} $  with the minimal GPU count such that its computation fully overlaps with the critical section and introduces no stalls.
\end{enumerate}

This two-stage decomposition offers two key benefits. First, it dramatically reduces the complexity of solving the original joint optimization problem. By decoupling the search into a sequence of smaller subproblems, the combinatorial explosion across sections is avoided. Once the near-optimal configuration for the critical section is identified, each auxiliary section can be configured independently and efficiently, without revisiting the global search space. Second, the approach achieves near-optimal end-to-end throughput. Because pipeline performance is fundamentally bounded by the critical path, maximizing the efficiency of the critical section brings the system close to its theoretical throughput limit. Resources allocated to auxiliary sections are kept to the minimum necessary to avoid stalls, ensuring that any overhead remains negligible. As a result, our method delivers high system efficiency while remaining computationally tractable.

\subsection{Section Communication}

\begin{figure}[!htbp]
\begin{center}
\centerline{\includegraphics[width=0.7\columnwidth]{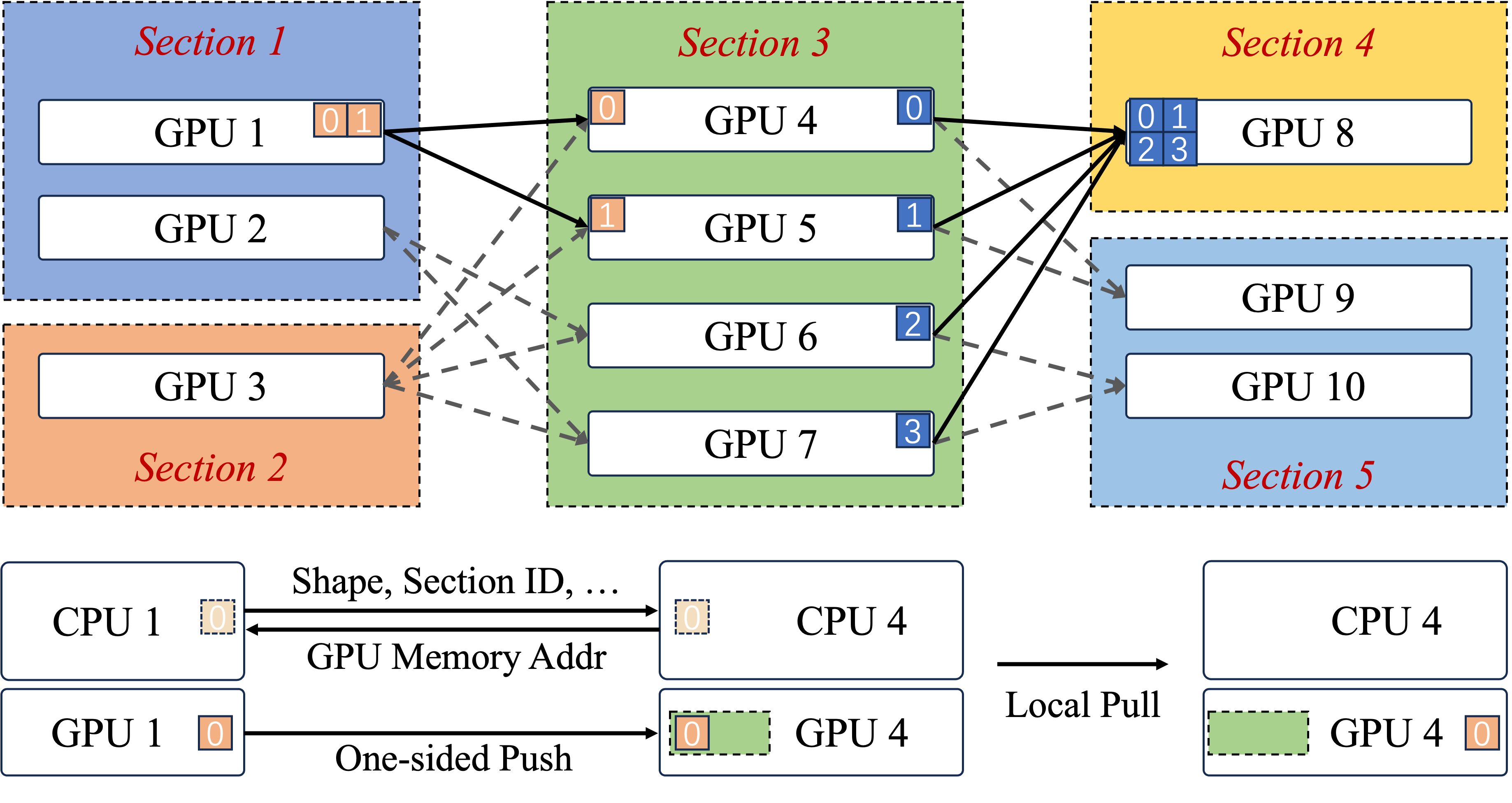}}
\caption{Illustration of data resharding and communication across sections.}
\label{fig:section_data_reshard}
\end{center}
\end{figure}

Following the steps above, managing tensor transfers across sections becomes essential. Since each section may adopt heterogeneous parallelism configurations—particularly in tensor parallelism (TP) and context parallelism (CP)—along with distinct \emph{fan-out} mechanisms, correctly and efficiently handling data resharding is critical. Moreover, as section executions are highly asynchronous with respect to one another, the communication process must not interfere with GPU computation and should achieve effective overlap with kernel execution to maximize hardware utilization.

To satisfy this requirement, we design and implement a high-performance, asynchronous, asymmetric message queue that supports the M-to-N communication pattern—a novel paradigm distinct from conventional point-to-point or collective communication in LLM training scenarios—to accommodate heterogeneous TP/CP configurations across section boundaries.

During communication establishment, the message queue decouples M-to-N communication into multiple point-to-point channels. Each channel comprises a CPU subchannel for metadata (e.g., tensor shape, section name, position within the TP/CP group, etc.) and a GPU subchannel for tensor data; both employ one-sided operations to enable asynchronous communication with negligible CPU/GPU interference. At runtime, each sender invokes the \texttt{push} API to transmit a tensor to remote receivers. The MessageQueue first leverages the CPU subchannel to exchange metadata and reserve a slot in the destination device's memory; the sender GPU then pushes tensor data directly without awaiting the receiver's participation. On the receiving side, the \texttt{pull} API dequeues the earliest message. When multiple senders contribute to a single tensor, the API automatically gathers the sharded fragments. Finally, the receiver updates local memory statistics to facilitate subsequent requests.

With this MessageQueue, we can efficiently accommodate diverse cross-section communication patterns—such as fan-out communication among data-parallel ranks and M-to-N transfers across heterogeneous TP/CP parallelism domains.
% In our current implementation, within each data-parallel replica of the upstream section, the TP0CP0 rank first gathers all locally split data shards. It then sends the complete tensor to the TP0CP0 ranks of the corresponding data-parallel replicas in the downstream section. Finally, each downstream TP0CP0 rank scatters the received data across its local tensor–context parallel communication group. We currently implement the above point-to-point communication using the NCCL backend, and plan to migrate to a single-sided RDMA or NVSHMEM-based design (e.g., using remote direct memory access or shared memory) to achieve higher performance and lower communication overhead.

\subsection{Section Computation Scheduling}
%The key to fully realizing Maestro’s performance benefits lies in maximizing inter-section parallelism while minimizing idle time caused by cross-section data dependencies.
While section-tailored training strategies and fine-grained resource allocation form the foundation of Maestro's efficiency, the decisive factor in unlocking its full performance potential lies in wavefront scheduling—a mechanism that maximizes inter-section parallelism while minimizing idle time induced by cross-section data dependencies.
During each training iteration, given a global batch of samples and the resulting constructed sections, Maestro first analyzes the sections that each sample will activate. In principle, one could globally search for an optimal sample ordering that minimizes the aggregated idle time across all sections. However, this combinatorial optimization is computationally intractable for practical workloads. As we noted earlier, in most scenarios we have encountered, the critical section—the bottleneck stage that dominates the execution timeline—has a decisive impact on overall training efficiency. Thus, we design our scheduling policy to prioritize the critical section: it is never allowed to stall waiting for forward or backward passes from other sections. This ensures the critical section remains continuously saturated, thereby maximizing overall training throughput. 

Guided by the critical-section-first principle, a 6-tuple $Sample(t\_f\_bc, t\_f\_c, t\_f\_ac, t\_b\_bc, t\_b\_c, t\_b\_ac)$ is adopted by Maestro to model each sample, where
$t\_f\_bc$, $t\_f\_c$, and $t\_f\_ac$ denote the execution times before, within, and after the critical section during the forward pass, respectively;
$t\_b\_bc$, $t\_b\_c$, and $t\_b\_ac$ denote the corresponding times during the backward pass.
We implement a heuristic scheduling algorithm that produces a near-optimal schedule in $O(N^2)$ time complexity where N denotes the number of samples per DP rank. In practical training scenarios with data parallelism, each DP rank typically processes tens to hundreds of samples. Moreover, since the scheduling process could fully overlaps with GPU execution, the $O(N^2)$ complexity introduces negligible overhead to end-to-end training latency.
% algorithm detail
In detail, the scheduling process operates in two phases as shown in Algorithm \ref{algo:section_scheduling}: First, it sorts all samples in ascending order of $t\_f\_bc$—prioritizing those that reach the critical section earliest—and initializes the result schedule with the top-ranked sample. Subsequently, for each remaining sample, the algorithm evaluates all feasible insertion positions within the partially constructed result schedule. For each candidate position, it simulates the end-to-end execution timeline across all sections to compute the resulting makespan (total execution time), with inter-section overlap explicitly modeled; the algorithm then commits the insertion that minimizes this metric.
This greedy insertion heuristic effectively approximates critical-path scheduling: by prioritizing samples with minimal $t\_f\_bc$ and dynamically selecting insertion points that suppress pipeline bubbles, it maintains continuous saturation of the critical section while opportunistically packing non-critical work into idle slots. The approach thus achieves near-optimal throughput without incurring the combinatorial cost of exhaustive search.

When the data parallelism degree exceeds 1, Maestro first partitions the global batch across DP ranks with the goal of balancing the distribution of activated sections. Each DP rank then independently executes the scheduling algorithm described above.
For sections with $\texttt{fanout}>1$, each DP rank merges the per-rank schedules from the $\texttt{fanout}$ DP ranks of its downstream section via round-robin interleaving. This ensures fair progression across parallel branches and prevents starvation of any downstream consumer, which is critical for maintaining balanced pipeline occupancy under heterogeneous data dependencies.

\begin{algorithm}[htbp]
    \caption{SectionSchedulingAlgorithm}
    \label{algo:section_scheduling}
\begin{algorithmic}[1]
    \State Given $Samples=List[(sample\_idx,t\_f\_bc, t\_f\_c, t\_f\_ac, t\_b\_bc, t\_b\_c, t\_b\_ac)]$.
    \State $InitialSampleOrder = SortByForwardTimeBeforeCriticalSection(Samples)$.
    \State $ResultOrder=[InitialSampleOrder[0]]$
    \For {$Sample \in InitialSampleOrder[1:]$}
        \State $best\_position, best\_makespan = 0, INF$
        \For {$position \in Range(0, length(ResultOrder) + 1)$}
            \State $current\_makespan = calculate\_makespan(ResultOrder, position, Sample)$
            \If {$current\_makespan < best\_makespan$}
                \State $best\_makespan, best\_position = current\_makespan, position$
            \EndIf
            \State $insert\_sample(ResultOrder, best\_position, Sample)$
        \EndFor
    \EndFor
\end{algorithmic} 
\end{algorithm}

\begin{figure}[!htbp]
\begin{center}
\centerline{\includegraphics[width=1\columnwidth]{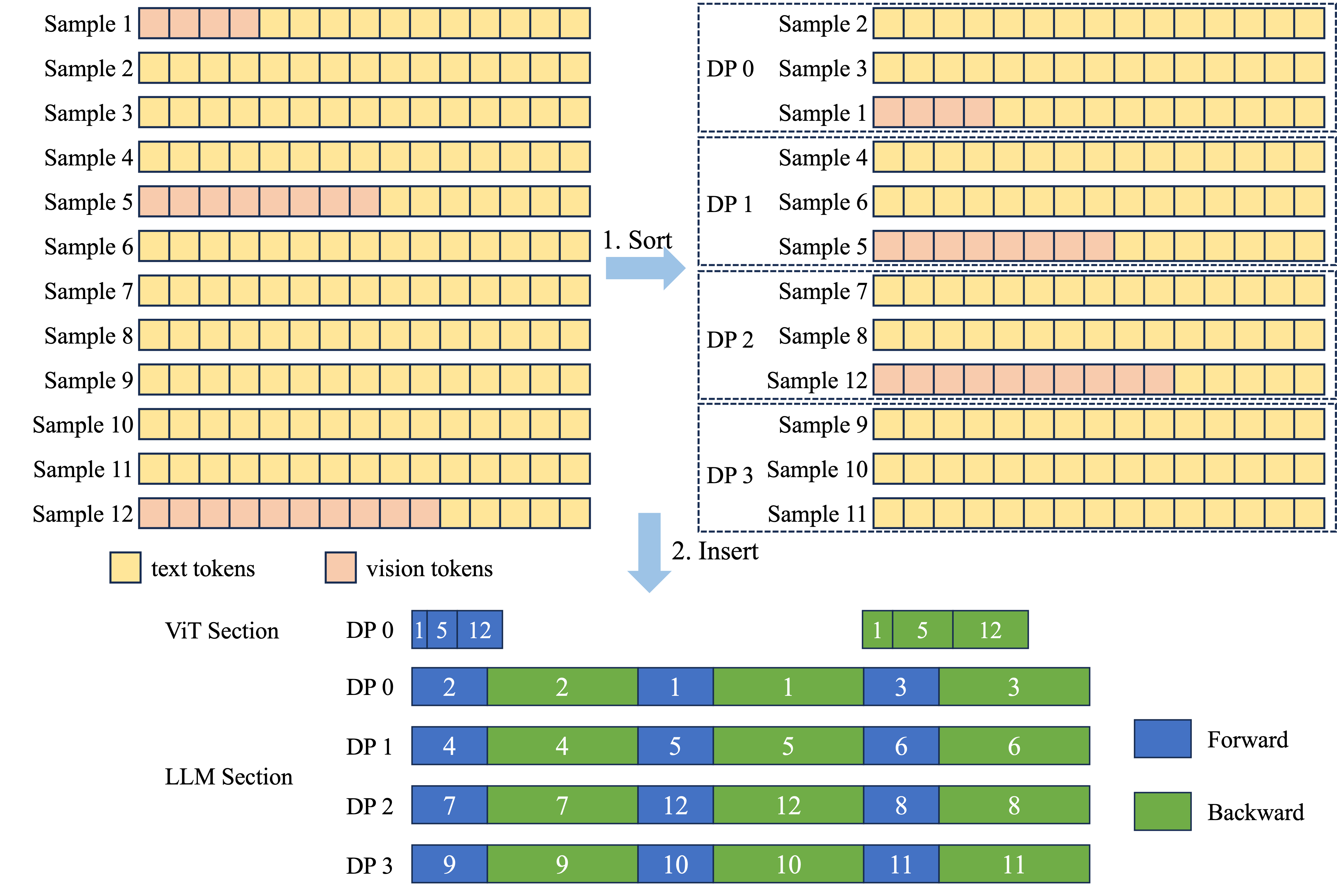}}
\caption{Data scheduling for VLM training with $\texttt{fanout}=4$ over a global batch of 12 samples.}
\label{fig:omni_data_schedule}
\end{center}
\end{figure}

We illustrate our approach using a vision-language model (VLM) training example, as shown in Figure \ref{fig:omni_data_schedule}. In this workload, two sections are constructed: the ViT section and the LLM section. The ViT section has a $\texttt{fanout}$ of 4, meaning each data-parallel replica in the ViT section supplies data to four replicas in the LLM section. With a global batch size of 12, each LLM replica processes 3 samples per iteration.
Clearly, the LLM section is the critical section in this scenario. The samples in their original order are represented in $Sample(t\_f\_bc, t\_f\_c, t\_f\_ac, t\_b\_bc, t\_b\_c, t\_b\_ac)$ as:
(0.1, 1, 0, 0, 2, 0.2), (0, 1, 0, 0, 2, 0), (0, 1, 0, 0, 2, 0), and (0.2, 1, 0, 0, 2, 0.4)....
Here, $t\_f\_bc$ and  $t\_b\_ac$ actually describe the forward and backward execution time in the ViT encoder respectively; a value of $t\_f\_bc = 0$ indicates a text-only sample that bypasses the ViT section and is fed directly into the LLM section. Since no additional modules follow the LLM backbone, all samples have $t\_f\_ac = 0$.
Applying Algorithm \ref{algo:section_scheduling} to samples on each DP rank and merging the resulting schedules at the ViT section yields the execution order shown in Figure \ref{fig:omni_data_schedule}. We can see that under this ordering, the LLM section remains continuously saturated—never stalling to await forward or backward passes from the ViT section. Consequently, ViT computation is fully overlapped with LLM execution, contributing zero critical-path overhead to end-to-end training throughput.

\section{Evaluations and Case Study}

We benchmark Maestro's efficacy on two representative production-scale workloads that embody distinct forms of heterogeneity:
(1) Multimodal training of the recently released Qwen3.5-400B-A17B and Qwen3-Next-80B-A3B models on a dataset with a 32K sequence length—exemplifying static heterogeneity (vision encoder vs. LLM backbone) and dynamic heterogeneity (text-only vs. text-image sample activation); and
(2) KL divergence-based knowledge distillation, where a frozen Qwen3.5-400B-A17B teacher guides the training of a Qwen3-Next-80B-A3B student~\citep{Qwen3_Next_80B_A3B_Instruct}—epitomizing cross-model execution asymmetry (forward-only teacher vs. full backpropagation student).
All experiments are conducted on Alibaba Cloud's large-scale GPU clusters. We adopt Megatron-LM~\citep{shoeybi2019megatron}—the foundation upon which Maestro is built—as our primary baseline to ensure a fair comparison under identical low-level optimizations. This baseline represents the state-of-the-art in homogeneous parallel training and serves as the de facto standard for large-scale LLM training systems. In all experiments, we ensure that the critical section's resource allocation remains identical to the baseline configuration to enable a fair comparison. Specifically, for VLM training where the baseline employs 512 GPUs, we allocate exactly 512 GPUs to Maestro's LLM section and provision additional resources exclusively for the ViT section. Similarly, in distillation experiments, the student section is allocated the same GPU count as the baseline, with extra resources dedicated to the teacher section.

We evaluate training performance using two primary metrics: (1) end-to-end token throughput, and (2) per-gpu token throughput—the latter isolates the efficiency gain attributable to Maestro's sectioning mechanism by factoring out scale effects.
In the VLM training experiments, we further measure relative efficiency against text-only training to quantify whether the performance overhead introduced by the ViT encoder on the LLM backbone has been fully eliminated. Notably, this methodology—comparing multimodal throughput to text-only baselines—has also been adopted by recent community efforts including Kimi-K2.5~\citep{Bai2026KimiKV} and LongCat-Flash-Omni~\citep{Team2025LongCatFlashOmniTR}.

\subsection{Vision-Language Model Training}
We first evaluate Maestro on training two recently released models, the Qwen3.5-400B-A17B and the Qwen3-Next-80B-A3B on a 32K-sequence lengthed multimodal dataset. Qwen3.5-400B-A17B is a native multimodal model, whereas for Qwen3-Next-80B-A3B we integrate the ViT encoder from Qwen3.5-400B-A17B to evalute multimodal training at this model scale.
The models contain two structurally and computationally asymmetric components: a ViT encoder for image feature extraction and an LLM backbone for cross-modal reasoning. In Maestro, these map naturally to two distinct logical sections—the ViT section and the LLM section. The ViT processes raw image patches into visual tokens, which undergo a 4:1 downsampling before concatenation with text tokens to form the complete input for the LLM backbone, as illustrated in Figure \ref{fig:vl_downsample}. Consequently, the ViT operates on a significantly longer token sequence than the effective visual context ultimately consumed by the LLM.
To accommodate this disparity, we employ section-specific parallelization strategies: the ViT section primarily leverages context parallelism to efficiently handle its long input sequences, while the LLM section adopts its own optimal parallelism configuration—tailored to its model scale and sequence characteristics. 
Besides, in practice, training VLM typically requires careful curation of the vision-to-text data mix. For instance, Kimi-K2.5~\citep{Bai2026KimiKV} adopts a 1:9 vision-to-text sample ratio, while LongCat-Flash-Omni~\citep{Team2025LongCatFlashOmniTR} uses a 1:2 ratio. This inherent modality skew creates a natural opportunity for cross-section computation overlap. With our wavefront scheduling mechanism, we can fully hide the computational cost of the ViT section—effectively eliminating its contribution to the end-to-end training time.

\begin{figure}[!htbp]
    \centering
    \includegraphics[width=0.8\linewidth]{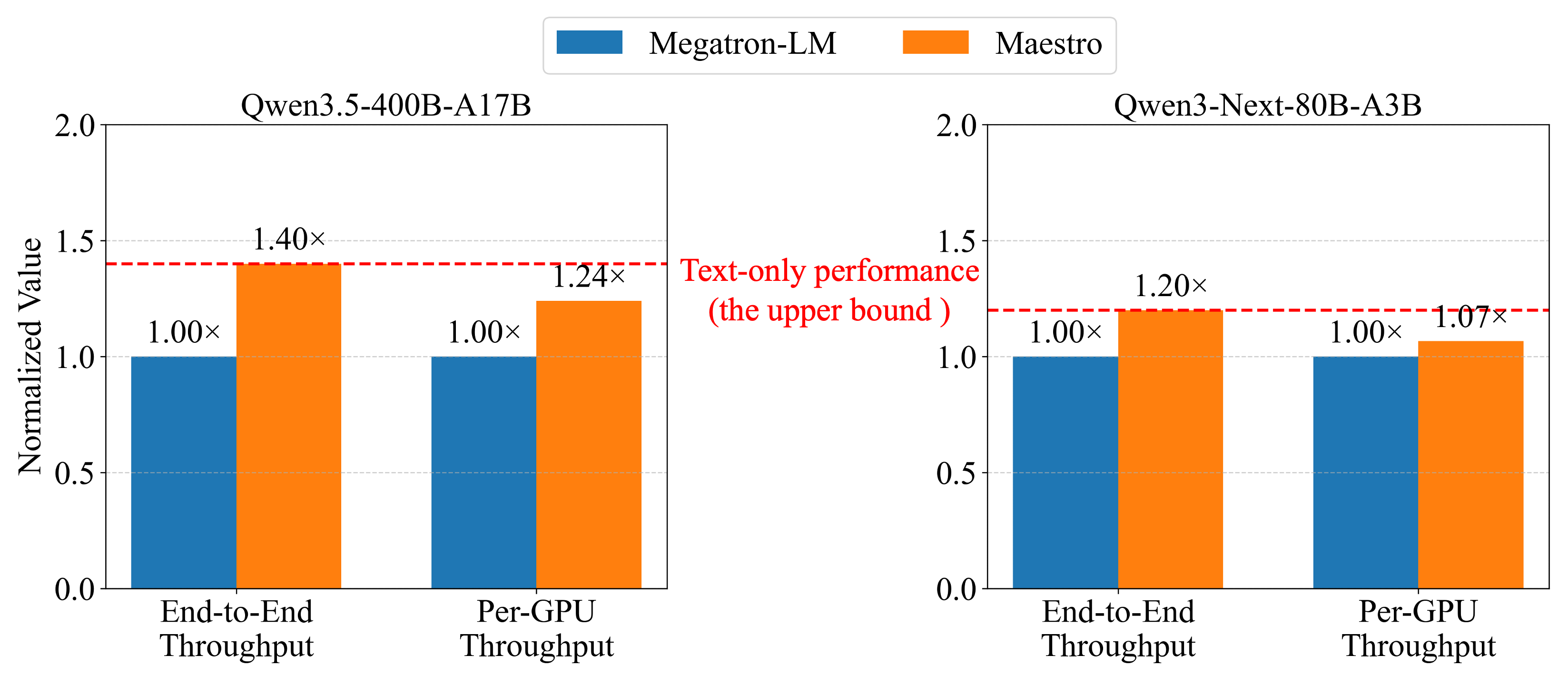}
    \caption{Maestro performance on multimodal training.}
    \label{fig:maestro_performance_multimodal}
\end{figure}

In both experiments, the LLM backbone uses the best available parallelism strategy and allocates an additional 12.5\% of resources to the ViT section. Specifically, Qwen3.5-400B-A17B is trained with a pipeline parallel size of 4, whereas Qwen3-Next-80B-A3B is trained without pipeline parallelism.
As shown in Figure \ref{fig:maestro_performance_multimodal}, Maestro achieves approximately 1.4× higher end-to-end training throughput than Megatron-LM on Qwen3.5-400B-A17B and 1.20× higher throughput on Qwen3-Next-80B-A3B.
Moreover, even when accounting for the additional resources allocated to the ViT section, Maestro still delivers 1.24× and 1.067× higher per-gpu throughput, respectively—demonstrating improved resource utilization. Finally, both experiments achieve 100\% relative training efficiency with regard to end-to-end throughput compared to text-only training, confirming that the performance overhead of the ViT section has been completely eliminated.

% Critically, during multimodal training, batches contain both multimodal (image–text) and pure-text samples in a fixed ratio \citep{zhu2025internvl3, wang2025internvl3}. This means the ViT section is only activated for multimodal samples and remains idle for pure-text ones. 

% 数据配比介

%To exploit the sparsity in ViT section activation, our scheduler overlaps the execution of the ViT section on multimodal samples with the LLM section’s processing of pure-text micro-batches, as shown in Figure \ref{fig:vlm_schedule}. Specifically, when a multimodal sample enters the pipeline, its ViT computation is launched concurrently with the LLM’s forward pass on a pure-text sample. This concurrency prevents the LLM section from stalling while waiting for ViT completion. As a result, the LLM maintains high utilization even when ViT sections are active, effectively hiding ViT latency within the natural concurrency of heterogeneous data batches.

\subsection{Distillation}
We further evaluate a knowledge distillation scenario where Qwen3.5-400B-A17B serves as the teacher model and distills knowledge to the Qwen3-Next-80B-A3B student model via KL divergence loss. This training paradigm requires the frozen teacher to transmit logits to the student during the forward pass. Given the stark execution asymmetry—forward-only teachers versus full backpropagation for the student—one could naturally partition the workload into distinct sections: one section per teacher model and a separate section for the student.
However, as noted earlier, the logits tensor can be orders of magnitude larger than the hidden state from which it is computed in the teacher model's output layer. For instance, in Qwen3.5-series models~\citep{}, the vocabulary size is 250K while the hidden dimension is merely 4K—rendering the logits tensor 62.5× larger in size. To circumvent the prohibitive inter-section communication overhead of transferring such voluminous tensors, we colocate the teacher's final output layer with the student model within the same logical section, enabling compact hidden-state transfer instead.

% In distillation training of Qwen3.5 models, multiple teacher models provide logits tensors to guide the training of a single student LLM. Critically, teacher models are used in inference-only mode, which perform forward passes to generate targets but do not participate in backward computation or parameter updates. This asymmetry in execution semantics naturally motivates a separation of concerns: we assign each full teacher model to its own dedicated teacher section, while the trainable student model resides in a separate student section. This partitioning enables full inter-section parallelism, allowing teachers and the student to execute concurrently without synchronization during the forward pass.

% However, a naive section boundary, where each teacher outputs full logits of shape $\left[L,V\right]$, would incur prohibitive inter-section communication overhead. The vocabulary size $V$ (e.g., 250K in Qwen3.5) is orders of magnitude larger than the hidden dimension $H$ (e.g., 4k), making such an approach bandwidth-prohibitive. To eliminate this bottleneck, we co-locate the teacher LM head layers with the student model inside the student section. Consequently, each teacher section only transmits compact hidden states $h_t\in \mathbb{R}^{L \times H}$ to the student section, where logits are reconstructed locally. This reduces cross-section communication volume by up to ~60× (250k/40k) and alleviates bandwidth pressure.

\begin{figure}[!htbp]
    \centering
    \includegraphics[width=0.6\linewidth]{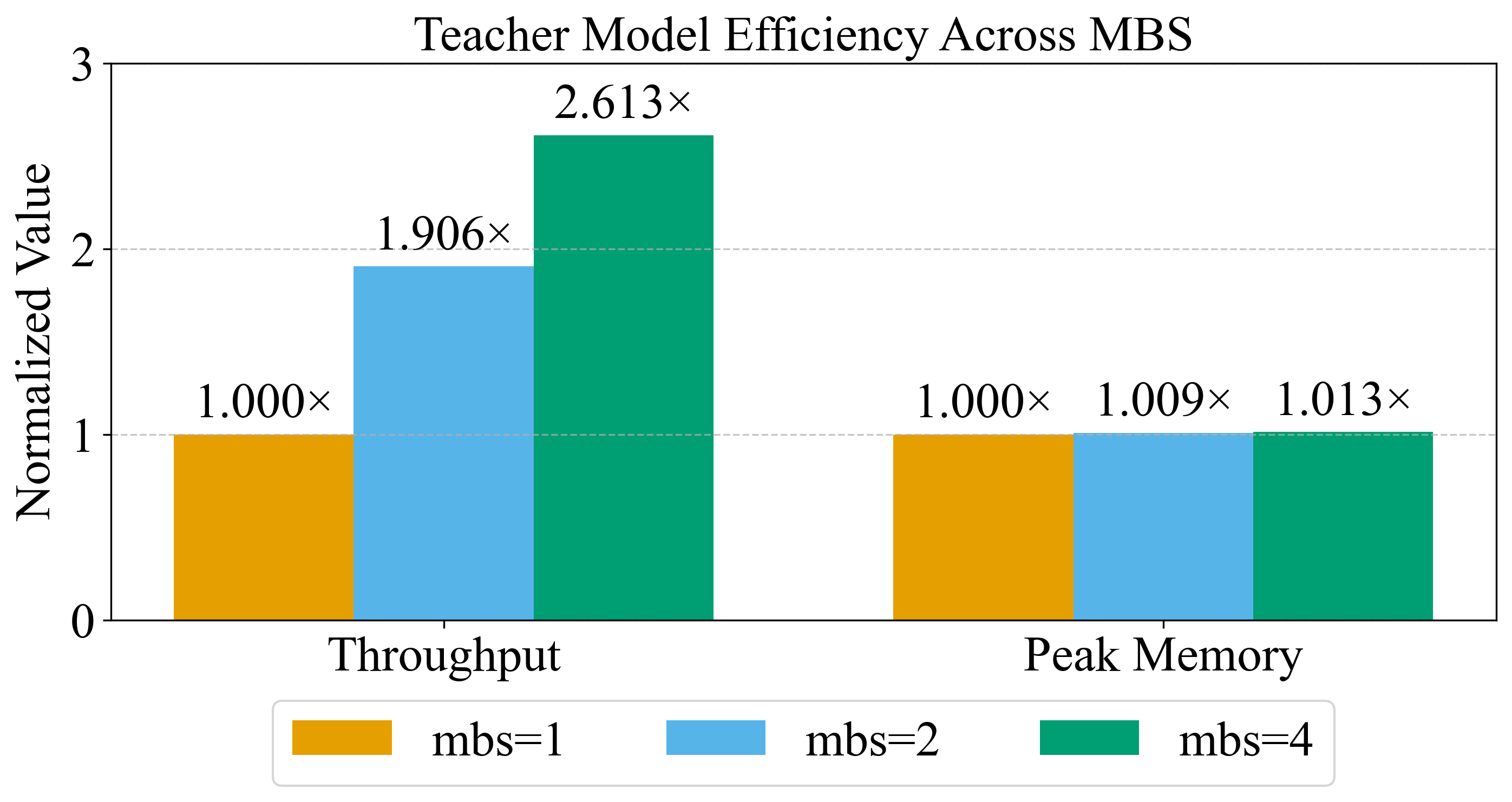}
    \caption{Normalized throughput and peak memory of the teacher model across micro-batch sizes.}
    \label{fig:distill_teacher_mbs}
\end{figure}

Furthermore, because the teacher section performs only forward passes (with frozen parameters), increasing its micro-batch size (MBS) substantially boosts throughput with negligible memory overhead. For instance, as shown in Figure \ref{fig:distill_teacher_mbs}, scaling the teacher MBS from 1 to 4 achieves a 2.6× throughput improvement while the peak memory consumption remains nearly flat. This insight enables fan-out execution: a single DP rank in the teacher section can concurrently serve multiple downstream DP ranks in the student section. Consequently, Maestro reduces the additional GPU allocation required for the teacher without compromising end-to-end throughput—effectively improving resource utilization.

\begin{figure}[!htbp]
    \centering
    \includegraphics[width=0.5\linewidth]{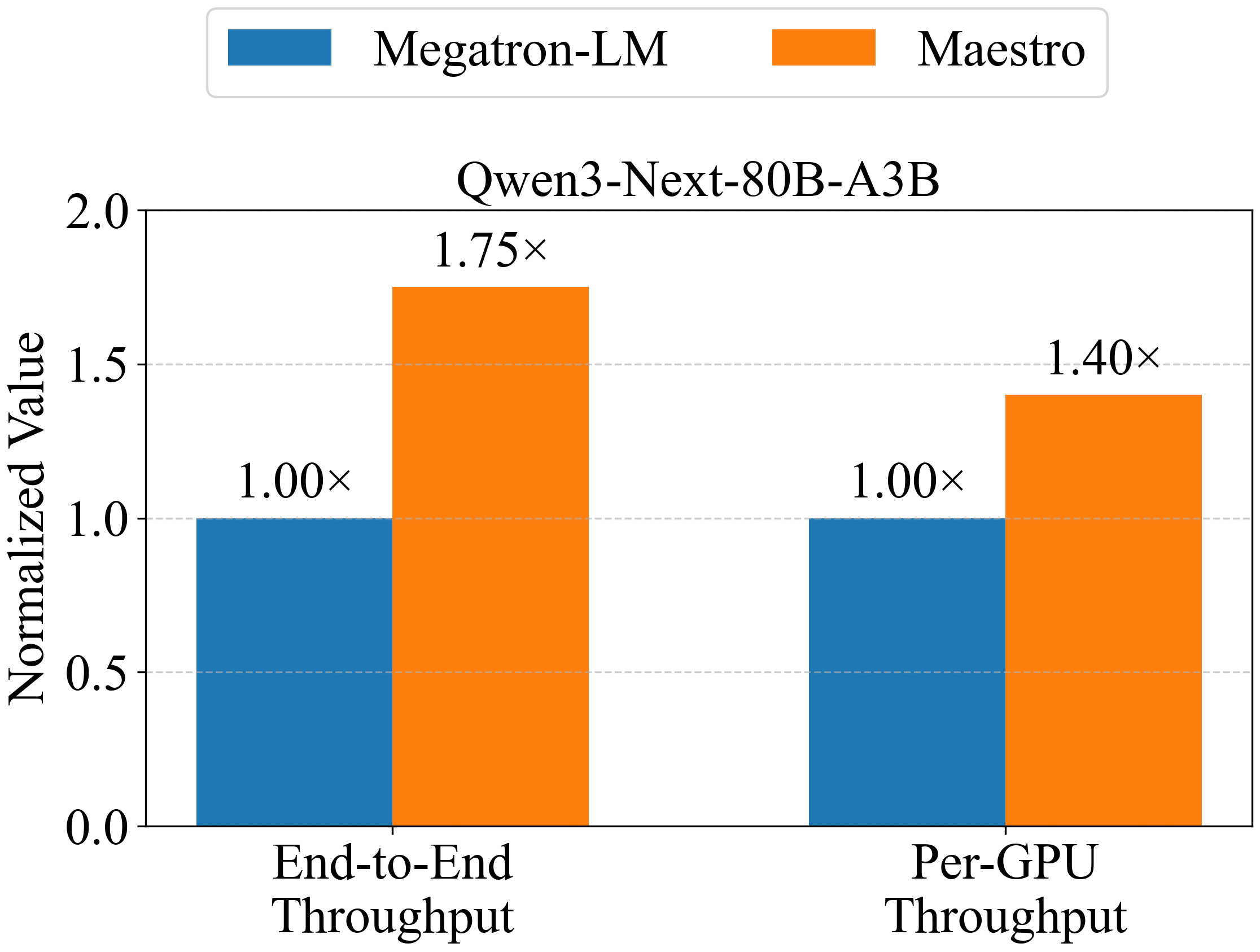}
    \caption{Maestro performance on distillation training.}
    \label{fig:maestro_performance_distillation}
\end{figure}

The experiments allocate an additional 25\% GPU resources to the teacher section relative to the baseline. As shown in Figure \ref{fig:maestro_performance_distillation}, Maestro achieves 1.75× higher end-to-end token throughput and 1.4× higher per-gpu token throughput compared to Megatron-LM—demonstrating simultaneous improvements in both absolute training performance and resource efficiency.
%Moreover, with the additional 25\% GPU resources allocated to the teacher section, Maestro achieves nearly 100\% relative training performance compared to the student-only baseline—indicating that the teacher section's computation is fully overlapped and contributes zero critical-path overhead to end-to-end token throughput.

\section{Related Work}
\subsection{LLM Training}
\textbf{Data Parallelism. }In data parallelism~\citep{li2020pytorch}, the entire model, including all parameters, is replicated identically across all participating devices. During training, the global mini-batch is partitioned into equal-sized sub-batches, with each device processing one sub-batch independently through the full forward and backward passes. After the backward pass completes on each device, the computed gradients for all parameters are synchronized across the device group using collective operations, ensuring that every replica accumulates the sum of gradients over the entire mini-batch. The optimizer then applies identical parameter updates on all devices, maintaining model consistency. Based on this, ZeRO~\citep{rajbhandari2020zero} further partitions optimizer states, gradients, and model parameters within the data-parallel group to reduce per-rank memory footprint.

\textbf{Tensor Parallelism. }Tensor parallelism (TP)~\citep{shoeybi2019megatron, chang2024flux}, also known as model parallelism, partitions individual model layers (e.g., attention heads or feed-forward networks) across multiple devices. For instance, the weight matrices of linear layers are split column-wise or row-wise, and intermediate activations are communicated via All-Gather or Reduce-Scatter operations. TP reduces per-device memory usage and enables training of extremely wide layers, but introduces fine-grained communication between devices at every layer, which can become a bottleneck on low-bandwidth interconnects.

\textbf{Pipeline Parallelism.}
Pipeline parallelism (PP)~\citep{huang2019gpipe, narayanan2019pipedream} partitions a model vertically into multiple sequential stages, each assigned to a dedicated device or group of devices. To overlap computation across stages, the global mini-batch is divided into smaller micro-batches that flow through the pipeline in a staggered manner—enabling concurrent execution and improving hardware utilization. A line of work~\citep{huang2019gpipe, narayanan2019pipedream, narayanan2021efficient, qi2024zero} has focused on designing advanced scheduling strategies that optimize throughput, memory footprint, and pipeline occupancy. 

\textbf{Expert Parallelism.}
Expert parallelism (EP)~\citep{singh2023hybrid,cai2024shortcut} is specialized for sparse mixture-of-experts architectures, where each input token is routed to a small subset of expert subnetworks out of a large pool. In EP, experts are distributed across devices, and each device hosts a disjoint subset of experts. During the forward pass, tokens are dispatched to the devices hosting their selected experts; after expert computation, results are gathered back to the originating device. EP minimizes redundant computation and memory footprint by ensuring that only activated experts are executed and stored, but incurs dynamic communication costs due to data-dependent routing. It is often combined with DP and TP to scale MoE models efficiently.

\subsection{MLLM Training}
Multimodal training extends LLM training by incorporating additional, typically smaller modality encoders. This introduces model heterogeneity: the encoder and LLM decoder exhibit vastly different compute and memory demands, necessitating distinct parallelization strategies for each component. Moreover, multimodal inputs bring dynamic data heterogeneity—such as variable sequence lengths in visual tokens—which causes fluctuating per-sample computation times. In pipeline parallelism, this variability leads to dynamic, data-dependent pipeline bubbles, significantly degrading throughput and complicating scheduling.
Optimus~\citep{feng2025optimus} addresses model heterogeneity by prioritizing the LLM backbone: it allocates GPU resources to the LLM first and then fits the encoders into the remaining memory budget. This design inherently bottlenecks encoder throughput, as they must operate under a severely constrained memory budget dictated by the LLM’s dominant footprint.
Recently, DistTrain~\citep{zhong2025disttrain} proposes inter- and intra-microbatch reordering algorithms to balance computation across data-parallel groups and reduce pipeline bubbles caused by variable visual data lengths, respectively. However, its focus is limited to alleviating imbalance within pipeline parallelism by minimizing PP bubbles. Similarly, PipeWeaver~\citep{chen2025pipeweaver} addresses data-length dynamicity by searching for pipeline schedules tailored to the current training batch.

\section{Conclusion}
\label{conclusion}

As model architectures evolve from monolithic layer stacks to composites of heterogeneous functional components, the foundational assumption of workload uniformity—upon which conventional training frameworks are built—is violated. This breakdown leads to suboptimal training efficiency and underutilized resources. The problem is further exacerbated when these components are data-dependent activated, as runtime irregularities amplify performance degradation.
To address this emerging class of compound workloads, we propose Maestro. Maestro decomposes the original computation graph into distinct sections, enabling each section to adopt a tailored training strategy for fine-grained resource allocation. To mitigate runtime irregularities caused by sparse section activation, we further introduce a wavefront scheduling algorithm that maximizes cross-section parallelism and minimizes stall time due to inter-section data dependencies. Experiments on compound workloads—such as native multimodal training and knowledge distillation—demonstrate that Maestro significantly improves both cluster-wide training throughput and per-GPU utilization. In the VLM training scenario, Maestro achieves 100\% relative performance compared to text-only training, fully eliminating the performance overhead of the additional ViT module through section computation scheduling. We share the design and implementation of Maestro in the hope of empowering the community to run next-generation compound LLM workloads with optimal performance.

% This conclusion is strong and hits all the key industrial track points: "production," "solves... problems," "delivering 1.48x," "delivering over 20% gains," and "industrial-scale."
%This paper introduces Maestro, a unified framework from our production environment that solves the two primary challenges in modern heterogeneous training: static and dynamic heterogeneity. Maestro replaces the monolithic paradigm with a flexible two-layer system. The \modelsection{} abstraction provides static disaggregation of parallelism and data policies, crucial for workloads like distillation, delivering a 1.48$\times$ speedup[cite: 131]. The \maestrotask{} abstraction provides dynamic, data-aware scheduling, critical for sparsely-activated multi-modal models, delivering an additional 1.2$\times$ (20\%+) gain[cite: 7]. By addressing both problems, Maestro provides a robust, general, and high-performance solution for the next generation of industrial-scale AI training.

\bibliography{maestro}
\bibliographystyle{colm2024_conference}

\end{document}